\documentclass[doublecol]{epl2}   

\title{Electromagnetic Vortex solitons in the magnetospheres of super-Eddington AGN}
\shorttitle{Title} 

\author{V.I. Berezhiani\inst{1,2} \and Z. Osmanov\inst{1} }

\institute{                    
  \inst{1} School of Physics, Free university of Tbilisi, Tbilisi 0159, Georgia\\
  \inst{2} Andronikashvili Institute of Physics (TSU), Tbilisi \ 0177, Georgia
}

\abstract{
Propagation and dynamics of electromagnetic vortex solitons have been studied in the electron positron ion plasmas of super-Eddington active galactic nuclei (AGN). Existence and stability of such structures is demonstrated for highly transparent AGN plasma media. Possible implications of dynamical features and corresponding observational signatures are discussed.}

\begin{document}

\maketitle

\section{Introduction}
 
Many astrophysical objects are characterised by high energy flux of electromagnetic (EM) radiation, the interaction of which with plasmas might lead to several interesting phenomena. Of significant interest is the phenomenon of twisted emission, because it is strongly believed that it might have several astrophysical applications \cite{harwit}. In particular, the author has examined the problem in the context of: astrophysical masers as probes for inhomogeneities in the interstellar medium; pulsars and quasars the SETI problems and the Kerr Black holes influencing the EM radiation. The latter has a very significant astrophysical application because by means of the radiation influenced by rotation, one can measure the spin of black holes. This method has been applied for the well known active galactic nuclei (AGN) M87 \cite{m87} and the authors concluded with high confidence that the supermassive black hole rotates with $\left(90\pm 5\right)\%$ of the maximum possible rate. In this regard it is important to note that in \cite{elias} the author has considered photon angular momentum in astronomy and defined corresponding observables. One should also note that photons with angular momentum might be produced either by the inverse Thomson scattering \cite{thom} or turbulence \cite{turb}, implying that the effects of angular momentum might take place in a variety of objects.

Generally speaking, EM radiation characterised by the photon angular momentum might influence either interstellar or intergalactic media. On the other hand, under certain conditions, transparent astrophysical plasmas with multi species of components may support electromagnetic high frequency soliton-like solutions for non-degenerate \cite{mikh,mach} as well as for highly degenerate media \cite{guga,BST,eliasson}, exhibiting other interesting phenomena \cite{rati}.

It is well known that nearby media of pulsars, magnetars and AGN are composed of relativistic electrons, positrons and ions \cite{carroll}. In \cite{vazha1,vazha3} the authors have considered the electron-positron plasma with a small fraction of ions. It was shown that large-amplitude, localised, electromagnetic spatiotemporal solitons are supported. In the light of the new finding \cite{harwit}, it is important to examine dynamics of twisted electromagnetic emission (vortices) in such plasmas. 

It is observationally evident that most of the astrophysical outflows like galactic and extragalactic jets or winds exhibit inhomogeneous internal structures \cite{carroll}. Normally the outflows are accompanied by EM radiation in most of the cases generated by the same structures. A special interest deserve the extragalactic jets emanating "from" the central cores of AGN - supermassive black holes, which, as we have already discussed, strongly influence emission characteristics.

In the present paper we consider the electron-positron-ion plasma in the magnetosphere of luminous AGN and we study the non-linear dynamics of the twisted EM pulse. We will show that under well defined conditions high power radiation can be trapped in a self-guided regime of propagation with one of the important observational signatures: zero field "cavities" in the center of the structures. As it will be argued, the structures will split into several sub-structures  - therefore, detection of multiple filaments surrounding the zero-field points at the center of the structure is regarded as an observational evidence of non-linear behaviour. We would like to emphasise that self-guiding phenomena allows pulse to propagate on long distances compared to linear diffraction length-scales allowing the pulse to maintain high intensity for large length-scales. Moreover, for relativistically strong EM radiation, the absorption legnth-scale increases considerably due to the plasma particle collisional absorption. In contrast to linear regime, the scale of propagation might enable us to detect them. The considered phenomena of  twisted pulse dynamics is a direct consequence of high efficiency of electron-positron pair creation. 

The paper is organized in the following way: in the next section we consider the theory of solitons, applying them to typical media of super-Eddington AGN and derive the governing equation, in the following section we discuss and obtain the results and  in the last section we shortly outline them.

\section{Main Considerations}

In this section we describe the propagation of electromagnetic waves in an electron-positron-ion plasma in the AGN magnetosphere with a small fraction of ions. Generally speaking, it is known that one of the necessary conditions of pair cascading $\gamma+\gamma -> e^++e^-$ by means of interaction of a relatively high energy ($\epsilon_{ph}$) photon with a soft photon is $\epsilon_{ph}>2m_ec^2$, where $m_e$ is the electron's mass and $c$ is the speed of light. As it is clear, the pair production might take place if $\epsilon_{ph}>1$MeV, which is easily satisfied in many AGNs. Another important condition is that the compactness parameter \cite{compact}
$$l = \frac{L\sigma_T}{Rm_ec^3} = 2\pi\times\frac{m_p}{m_e}\times\frac{L}{L_{Edd}}\times\frac{R_S}{R} \simeq$$ 
\begin{equation}
\label{l} 
\simeq 1.2 \times 10^3\times\frac{L}{L_{Edd}}\times\frac{10R_S}{R}
\end{equation}
should satisfy the condition $l>>10$. Here $m_p$ denotes the proton's mass, $L$ and $L_{Edd} = 4\pi GMcm_p/\sigma_T$ are respectively the luminosity and the Eddington limit of the AGN, $G$ is the gravitational constant, $\sigma_T$ denotes the Thomson cross section, $R_S = 2GM/c^2$ is the Schwarzschild radius of a supermassive black hole, $M$ is its mass and $R$ is a characteristic length-scale of the source. We normalise the length-scale by the typical value for accretion models \cite{carroll}. As it is evident from the aforementioned expression, in many AGNs (especially in super Eddington objects), the pair production process might be extremely efficient and therefore the number density of pairs will be much higher than the density of ions.

The disk accretion models around black holes allow the following expression for the disk temperature \cite{carroll}
\begin{equation}
\label{T1} T = \left(\frac{3GM\dot{M}}{8\pi\sigma R_S^3}\right)^{1/4}\left(\frac{R_S}{R}\right)^{3/4}
\left(1-\sqrt{\frac{R_S}{R}}\right)^{1/4},
\end{equation}
where $\dot{M} = L/(c^2\eta)$ is the accretion rate, $\eta<1$ is a dimensionless parameter characterising energy conversion efficiency and $\sigma$ is the Stefan-Boltzmann constant. If one considers parameters typical for AGNs, from Eq. (\ref{T1}) one obtains
\begin{equation}
\label{T2} T \simeq 8\times 10^{4}\times\left(\frac{0.25}{\eta}\times\frac{L}{L_{Edd}}\times\frac{M}{M_8}\right)^{1/4}\times\left(\frac{R_{10}}{R}\right)^{3/4} (K),
\end{equation}
where $M_8 = M/(10^8M_{\odot})$ and $R_{10} = 10R_s$. We have taken into account that for the extreme case $\eta = 1/4$ \cite{carroll}. From this expression one concludes that the temperature is non-relativistic and therefore relativistic correction to electron's mass can be neglected.

As it has been shown in \cite{vazha1} the equation governing the propagation of EM waves in plasmas with low ion concentrations compared to pairs is given by the following non-linear Schr{\''o}dinger equation in the dimensionless form
\begin{equation}
\label{shr1} 
i\frac{\partial A}{\partial t}+\Delta_{\perp}A+F\left(|A|^2\right)A = 0,
\end{equation}
where $A$ is a slowly varying amplitude of the circularly polarised EM wave, $\Delta_{\perp} = \partial^2/\partial x^2+\partial^2/\partial y^2$, $A\rightarrow |e|A/m_ec^2$, $e$ is the electron's charge, $t\rightarrow \epsilon^2\omega_e^2/(16\omega)\;t$,  $r_{\perp}\rightarrow \epsilon\omega_e/(8^{1/2}c)\;r_{\perp}$ and $\omega_e = \left(4\pi e^2n_{oe}/m_e\right)^{1/2}$ is the Langmuir frequency of the plasmas, $\epsilon = n_{0i}/n_{0e}(<<1)$, $n_{0i}$ and $n_{0e}$ denote the densities of ions and electrons respectively. It has been assumed that longitudinal (in the direction of propagation) extent of the circularly polarised pulse is small compared to the transverse extent. In the same time the plasma is highly transparent ($\omega>>\omega_e$) and the effect related to the group velocity dispersion ($\sim\partial^2A/\partial\xi^2$) can be neglected, where $\omega$ is the frequency of the EM pulse, $\xi = z-\upsilon_gt$ is the comoving coordinate, $z$ is the coordinate of propagation of the wave, $\upsilon_g$ is its group velocity ($\upsilon_g\simeq c$) and the saturation function is given by
\begin{equation}
\label{satur} 
F\left(|A|^2\right) = 1-\frac{1}{\left(1+|A|^2\right)^2}.
\end{equation}

\begin{figure}
  \resizebox{\hsize}{!}{\includegraphics[angle=0]{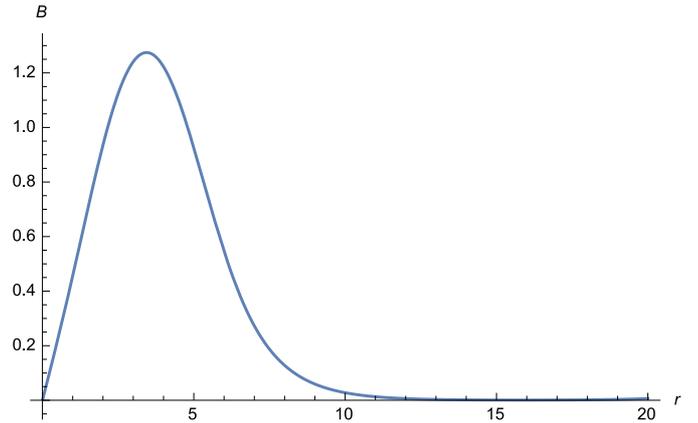}}
  \caption{Soliton solution of $B(r)$ for $m = 1$.}\label{fig1}
\end{figure}

\begin{figure}
  \resizebox{\hsize}{!}{\includegraphics[angle=0]{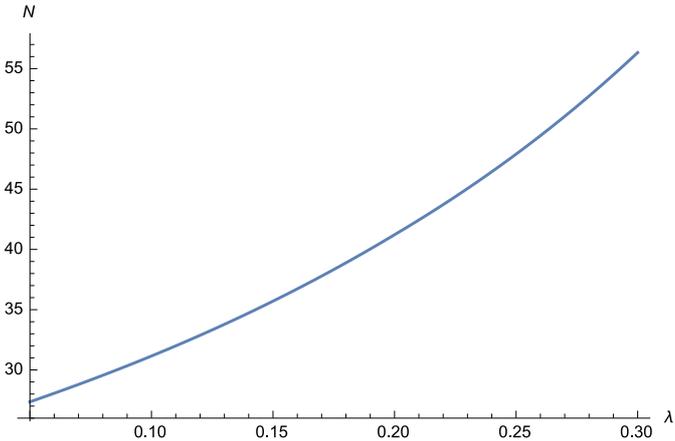}}
  \caption{The dependence of the photon number on $\lambda$.
The set of parameters is: $m = 1$ and $B(0) = 0$.}\label{fig2}
\end{figure}

\begin{figure}
  \resizebox{\hsize}{!}{\includegraphics[angle=0]{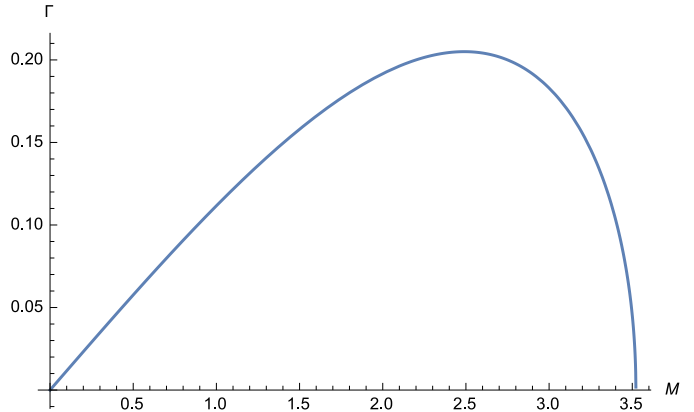}}
  \caption{Increment versus $M$.
The set of parameters is: $m = 1$, $B_m = 0.4$ and $r_{\star} = 5.5$.}\label{fig3}
\end{figure}

Here we study spatial solitons, therefore we rewrite Eq. (\ref{shr1}) in the cylindrical form by means of the anzatz
\begin{equation}
\label{anz} 
A = B(r)exp\left(i\lambda t+im\theta\right),
\end{equation}
obtaining
\begin{equation}
\label{shr2} 
\frac{d^2B}{dr^2}+\frac{1}{r}\frac{dB}{dr}-\lambda B-\frac{m^2}{r^2}B+F\left(B^2\right)B = 0,
\end{equation}
where $m$ is the integer number (topological charge), $\lambda$ is the non-linear frequency shift and $r = \sqrt{x^2+y^2}$.



For zero topological charge ($m=0$) Eq. (\ref{shr2}) has been solved for fundamental soliton solutions.  It was demonstrated that the solutions are stable against small perturbations \cite{vazha1,vazha3}. In contrast to these works we consider the case for non-zero topological charge corresponding to the vortex-soliton structures characterised by zero nodes at the center ($r = 0$). One can show that the aforemetioned equation has localised solutions provided that $0<\lambda<1$. The amplitude of the structure is an increasing function of $\lambda$. In Fig. 1 we plot the behaviour of $B(r)$ for $m = 1$ and $\lambda = 0.5$. As it is clear from the plot, the amplitude of $B$ becomes more than unity, indicating the relativistic regime.
In Fig. 2 we demonstrate the dependence of power $N = \int 2\pi r|B|^2dr$ versus $\lambda$. 

For fundamental solitons, the Vakhitov-Kolokolov criteria implies the following condition \cite{VK} 
\begin{equation}
\label{VK} 
\frac{\partial N}{\partial\lambda}>0.
\end{equation}
Unlike that case, for vortex-like solitons the mentioned condition guarantees stability only against radial perturbations.

On the other hand, in \cite{vincotte} it was found that against azimuthal perturbations the instability growth rate is given by
\begin{equation}
\label{GR} 
\Gamma = \frac{M}{r_{\star}}Re\left[\frac{4B_m^2}{(1+B_m^2)^3}-\frac{M^2}{r_{\star}^2}\right]^{1/2},
\end{equation}
where $B_m$ is the maximum value of $B$, $r_{\star}$ is the corresponding radial coordinate and $M$ is the integer number and represents the number of filaments the initial soliton will decay into. From numerical analysis of Eq. (\ref{GR}) one can straightforwardly show that for the topological number $m = 1$ the values of $B_m$ and $r_{\star}$ are respectively $0.4$ and $6$. In Fig. 3 we show the dependence of $\Gamma$ on M (for a moment we assume that $M$ varies continuously). The set of parameters is: $m = 1$, $B_m = 0.4$ and $r_{\star} = 5.5$. As it is clear, the growth rate becomes maximum for $M\simeq 2.5$, implying that the initial pulse will split into two or three solitons in total conserving the initial angular momentum, $mN$. 

We would like to emphasise that although the vortex-like solitons turn out to be unstable, however the growth rate of the instability drastically reduces for relativistically large amplitudes of the structure ($B_m>>1$, see Eq. (\ref{GR})) implying that the soliton might propagate a long distance almost maintaining its shape.

In general when twisted emission carrying angular momentum ($m\neq 0$) enters the AGN plasma media, it will  develop differently depending on the initial characteristics. As it has been demonstrated in \cite{firth} (see also \cite{MBM,skarka}) dynamical processes do not significantly depend on a type of the saturation function. Consequently we state that if the pulse parameters (power and amplitude) corresponding to a certain point in the area over the curve shown in Fig. 2 but closer to the curve itself, the pulse relaxes to the equilibrium vortex soliton solution. Subsequent vortex solitons undergo azimuthal decaying instability into fundamental soliton structures. If the power of the pulse is considerably above the critical value $N_c\simeq 25$, the complex behaviour of the pulse develops. Due to the modulation instability the pulse splits into filaments finally forming multiple interacting fundamental solitons. Though the structure of the pulse undergoes dramatic changes the zero field in the center of the vortices remains unaltered due to topological reasons \cite{vazha2}. This could be one of the important observational signatures, which might take place in the AGN media.


\section{Discussion and Results}
In the previous section we have found that the typical length scale is normalised by the value $\Lambda_{\perp}\simeq \sqrt{8}c/(e\omega_e)$, which leads to the following expression
$$\Lambda\simeq 170\times\left(\frac{\eta}{0.1}\times\frac{10^{-2}}{\epsilon}\times\frac{L_{Edd}}{L}\right)^{1/2}\times$$
\begin{equation}
\label{L} 
\times\left(\frac{R}{R_{10}}\right)^{3/4}\times\left(\frac{M}{M_8}\right)^{1/4} cm,
\end{equation}
where $R_{10}\equiv 10R_S$ and we have taken into account that the accretion number density is given by \cite{zev}
\begin{equation}
\label{dens} 
n_i\simeq\frac{L}{4\pi\eta m_pc^2R^2\upsilon_{acc}},
\end{equation}
where $m_p$ is the proton's mass, $\upsilon = (2GM/R)^{1/2}$ is the accretion velocity and $\eta$ represents efficiency of the accretion process. As we have already noticed, $r_{\star} = 5.5$, which means that in dimensionlal terms the overall length-scale after filamentation will be of the order of $\Lambda$. Therefore, typical length-scales of structures are small compared to $R_s$, which justifies our assumption of spatial coherence. From Eq. (\ref{dens}) one can straightforwardly show that the Langmuir frequency is of the order of $10^{10}$Hz. For guarantying plasma transparency we consider frequencies exceeding $\omega_e$. Then, for the power of critical EM radiation one obtains

$$P_c = N_c\times\frac{2m^2c^5}{\pi e^2\epsilon^2}\times\left(\frac{\omega}{\omega_e}\right)^2\simeq$$
\begin{equation}
\label{P} 
\simeq 1.4\times 10^{18}\times\frac{1}{\epsilon^2}\times\left(\frac{\omega}{\omega_e}\right)^2\; ergs\;s^{-1}.
\end{equation}
In this paper we consider cases with $\epsilon<<1$ and $\omega_e<<\omega$, therefore the critical power exceeds the value $1.4\times 10^{18}$ergs s$^{-1}$. Typical AGN emit in the broad energy band starting from radio to gamma rays. as an example it is easy to show that for radio frequency $10^{11}$Hz, and a parameter $\epsilon=10^{-2}$, the resulting power is of the order of $10^{24}$ ergs s$^{-1}$. On the other hand, the Eddington luminosity for the given mass of the considered supermassive black hole $M_8$ is $1.2\times 10^{46}$ ergs s$^{-1}$ and since we are interested in high density electron-positron pairs for the super-Eddington AGNs the luminosity will  be even higher. This means that the power of EM pulses in the radio band produced by AGNs will be much higher than the critical value. Therefore, as we have already discussed, by means of the modulation instability, the initial structure will split into multiple filaments, surrounding zero field at the center of the structure - an important fingerprint the emission pattern must be characterised with.

\section{Summary}
We have considered super-Eddington AGNs and studied the behaviour of solitary structures carrying angular momentum provided by the metric of Kerr black holes located in the central regions of active galaxies.

For the vortex-like solitons with the topological number $m = 1$ it has been demonstrated that the solitons are  radially stable, undergoing instability against azimuthal perturbations. However, for relativistically large amplitudes the increment of the azimuthal instability significantly reduces allowing the structure to maintain for long distances. In due course of propagation the structure will split into two or three fundamental solitons flying off tangentially to the solitary ring structures.

We have estimated the critical power which can be much less than typical luminosities provided by AGNs in the broad band emission spectra. Therefore, the pulses first split into filaments and after that form fundamental solitons leaving in the center zero field. 

\acknowledgments
The research of ZO was supported by the Shota Rustaveli National Science Foundation grant (NFR17-587).

\end{document}